\documentclass[a4paper]{article}

\usepackage{spconf}
\usepackage{amsmath,amsfonts,amssymb}
\usepackage{algorithm,algorithmic}
\usepackage{xcolor,graphicx}
\usepackage{bbm}
\usepackage{enumitem}
\usepackage{multirow} 
\DeclareMathOperator{\E}{\mathbb{E}}
\ninept

\setlist{nosep}

\title{Federated Self-Learning with Weak Supervision for Speech Recognition}
\name{ 
\parbox{\textwidth}{\centering{
  Milind Rao \quad Gopinath Chennupati \quad Gautam Tiwari \quad Anit Kumar Sahu \\ Anirudh Raju \quad Ariya Rastrow \quad Jasha Droppo }  }  \vspace{-0.2cm}}
\address{
Amazon Alexa AI, U.S.A.
  \vspace{-0.3cm}}

\begin{document}

\maketitle
\begin{abstract}
  Automatic speech recognition (ASR) models with low-footprint are increasingly being deployed on edge devices for conversational agents, which enhances privacy.
  We study the problem of federated continual incremental learning for recurrent neural network-transducer (RNN-T) ASR models in the privacy-enhancing scheme of learning on-device, without access to ground truth human transcripts or machine transcriptions from a stronger ASR model. In particular, we study the performance of a self-learning based scheme, with a paired teacher model updated through an exponential moving average of ASR models.
  Further, we propose using possibly noisy weak-supervision signals such as feedback scores and natural language understanding semantics determined from user behavior across multiple turns in a session of interactions with the conversational agent. These signals are leveraged in a multi-task policy-gradient training approach to improve the performance of self-learning for ASR. 
  Finally, we show how catastrophic forgetting can be mitigated by combining on-device learning with a memory-replay approach using selected historical datasets. These innovations allow for $10\%$ relative improvement in WER on new use cases with minimal degradation on other test sets in the absence of strong-supervision signals such as ground-truth transcriptions.
  
\end{abstract}

\vspace{-3mm}
\section{Introduction}
\label{sec:intro}
\vspace{-3mm}

On-device deployment of voice technologies enables use of conversational agents in settings without a reliable network connection to the cloud. It enables lower-latency responses by removing the need for utterances to be transmitted to the cloud for processing. Offline use, vehicular control, and healthcare are new use cases within this paradigm. When ASR is deployed on-device, models need to be adapted for specific acoustic or linguistic content specific to the deployment as well as temporal adaptation to distribution shifts in use across time. In this work, we look at continually and incrementally updating ASR models with resource constraints of memory and compute at the device in \emph{federated} settings, i.e., privacy-enhancing features where (1) utterances are not transmitted to the cloud, (2) persistent storage of audio is not required, and (3) human ground-truth annotations of the audio need not be obtained.  

\noindent Privacy-preserving machine learning~\cite{al2019privacy} can enable learning from user data while mitigating privacy risks. Federated learning (FL)~\cite{mcmahan2017communication} is one of the most popular privacy-preserving learning frameworks which involves training models on-device, with data not leaving edge devices. 
In FL, multiple model updates from a number of participating devices are aggregated securely on a central server at every round. FL has been demonstrated to perform well in speech applications such as speech recognition \cite{guliani2021training}, keyword spotting \cite{hard2020training}, and speaker verification \cite{granqvist2020improving} among others. Mixed centralized and federated training was done in \cite{hard2022production} and layer-wise representation learning in \cite{huo2022incremental}. However, the aforementioned works involve training a model from scratch instead of fine-tuning a well-trained model. In addition, previous works considered static data which does not change across rounds. Differently from previous work, we consider FL settings, where the model is initialized to a well-trained model and streaming data on devices, which are not persisted across rounds. In \cite{jia2022federated}, authors look at domain adaptation of ASR in a federated setting. We additionally look at incorporating weak supervision to learn from alternate sources of feedback. 

\noindent Semi-supervised learning (SSL) deals with training and improving ASR using unlabelled audio, such as the audio available at devices. Unsupervised approaches such as data2vec \cite{baevski2022data2vec} or WavLM \cite{chen2021wavlm} use contrastive objective functions to pretrain speech models that are then finetuned. Alternatively, a common paradigm is to use a stronger teacher model to label unlabelled data \cite{parthasarathi2019lessons} however this approach cannot be applied to the resource constrained setting of on-device learning. Noisy student learning or iterative pseudo-labelling approaches \cite{xu2021self,chen2020semi} use the ASR model to self-label clean audio with the model trained to predict the same label with augmented version of the audio. Here the audio could be additionally filtered to include cases where the model does not have low confidence. We build off the work in \cite{manohar2021kaizen} where hybrid HMM-DNN and connectionist temporal classification (CTC) ASR models are updated using a paired teacher model updated using an exponential moving average of the student model. These methods have not been applied to recurrent neural network-transducer (RNN-T) ASR models \cite{graves2012sequence} that are streaming compatible and widely used across ASR applications.

\noindent We combine self-learning in this work with weak supervision. In conversational agents, users interact across multiple turns in a session. As shown in prior works \cite{ponnusamy2020feedback}, later interactions can be used to determine if a request has been correctly handled. If a user cancels or repeats their request, dissatisfaction is signalled. The semantics of the terminal request can be used as feedback for the initial request. Although this is not the ground truth transcription, we use such signals to update ASR models. Users can also be prompted for an explicit feedback signal as another example for a feedback score. We use the REINFORCE \cite{williams1992simple,vesely2013sequence,prabhavalkar2018minimum,rao2021mean} framework to update models using arbitrary rewards. 

\noindent \textbf{Contributions:} We look at incremental updates to ASR models using unlabelled audio on edge devices with federated, compute and memory constraints. We show on public and internal datasets that:
\begin{itemize}[leftmargin=*]
    \item Self-learning with a paired teacher model updated through exponential moving average of ASR can be used to improve the performance of RNN-T by $10\%$ on new use cases;
    \item Rehearsal training using historical datasets for generating model updates~(pseudo-devices) at the cloud mitigates catastrophic-forgetting \cite{kirkpatrick2017overcoming} on other test sets in self-training;
    \item Self-learning performance is improved by including weak supervision of NLU semantics or noisy feedback scores integrated through a policy-gradient approach.
\end{itemize}

\section{Methods}

\vspace{-2mm}
\subsection{RNN-T ASR model architecture}
\label{sec:rnnt}
\vspace{-2mm}

The RNN-T~\cite{graves2012sequence} architecture used for real-time speech recognition consists of a model that predicts the probability $P\left(\mathbf{y}|\mathbf{x}\right)$ of labels $\mathbf{y}=\left(y_{1},...,y_{U}\right)$ given acoustic features $\mathbf{x}=\left(x_{1},...,x_{T}\right)$. It comprises an encoder, a prediction network, and a joint network. The encoder is analogous to an acoustic model that takes a sequence of acoustic input features, and outputs encoded hidden representations. The prediction network corresponds to a language model that accepts the previous output label predictions, and maps them to corresponding hidden representations. The joint network is a feed forward network that takes both the encoder and prediction network hidden representations, and predicts the final output label probabilities with softmax normalization. A model with parameters $w$ produces $m$-hypotheses ($\textbf{y}_1,...\textbf{y}_m$) given input $\textbf{x}$ with probability $p_w\left(\textbf{y}_i|\textbf{x}\right)$.
\if{0}
\begin{figure}[htp]
	\centering
	\includegraphics[width=0.75\linewidth]{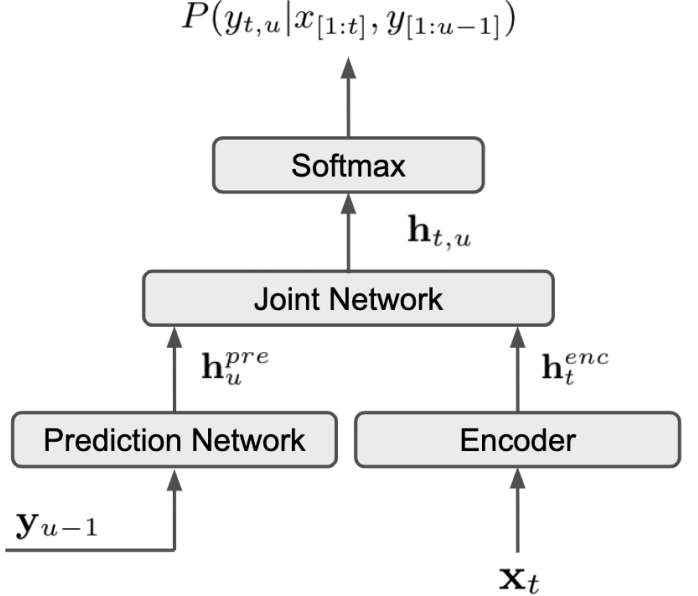}
	\caption{RNN-T ASR model architecture}
	\label{fig:rnnt} 
\end{figure}
\fi

\begin{figure}[t]
	\centering
	\includegraphics[width=0.95\linewidth]{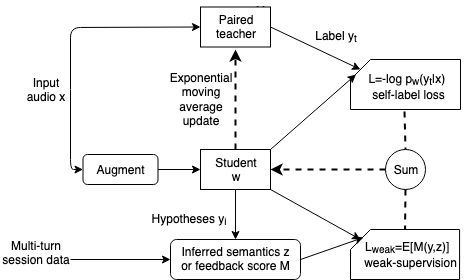}
	\caption{Workflow of federated self-learning with weak supervision: a paired teacher model produces a label with clean audio. The self-label loss enforces consistency for the student with augmented audio. Alternate weak supervision loss minimizes the expected feedback score for a hypothesis inferred from multi-turn session data.}
	\label{fig:self_learning_architecture} 
	\vspace{2mm}
\end{figure}

\vspace{-4mm}
\subsection{Federated Self-Learning for ASR}
\vspace{-2mm}

\begin{algorithm}[t]
	\caption{Federated self-learning for updating ASR with optional weak supervision including cloud pseudo-devices}\label{alg:self-learn}
	\begin{algorithmic}[1]
		\REQUIRE $N$ number of local steps per round, $\eta, \eta_k$ - global and on-device learning rates, $\delta$ exponential moving average rate (EMA), $u$ EMA update frequency, $\mathcal{L}\left(\cdot\right)$ RNN-T loss function optionally including $\mathcal{L}_{weak}\left(\cdot\right)$ weak supervision loss if such weak feedback is available on-device, $\mathcal{D}_{ht}$ data used for rehearsal training with ground-truth labels, $\mathcal{C}$ cloud pseudo-devices. 
		\ENSURE $w_\mathcal{G}^{r}$ incrementally updated global model, $w_\mathcal{T}^r$ updated teacher model
		\STATE Init. $w_\mathcal{G}^{0}$ \COMMENT{start training with a pre-trained model}
        \FOR{each round $r = 1, 2, \dots$}
        \STATE $\mathcal{S} \leftarrow$ (sample a subset of devices)
        \FOR{each device $k \in \mathcal{S} \cup \mathcal{C} $ \textbf{in parallel}}
        \STATE $w_{k}^{r} = w_\mathcal{G}^{r-1}$ \COMMENT{Global models broadcasted}
        \IF{$k \in \mathcal{S}$}
        \STATE  $\mathcal{D}_{train} \leftarrow$ (draw from $\mathcal{D}_{ssl}$ filtered utterances on device $k$ transcribed by teacher model $w^r_{\mathcal{T}}$, including weak supervision information $z$ and audio augmentation)
        \ELSIF{$k \in \mathcal{C}$}
        \STATE $\mathcal{D}_{train} \leftarrow$ (draw from $\mathcal{D}_{ht}$ for ASR rehearsal training on cloud pseudo-devices)
        \ENDIF
        \FOR{batch $b_{i}, \forall i \in \{1, \ldots, N\}$ from $\mathcal{D}_{train}$ }
        \STATE $w_{k}^{r} \leftarrow$ $\texttt{optimizer}_{k}$.\texttt{update}($\eta_k, \nabla \mathcal{L}(w_k^r; b_{i})$)
        \ENDFOR
        \STATE $\Delta w^r_k = w^r_k - w^{r-1}_{\mathcal{G}}$ \COMMENT{Transmit model delta to cloud}
        \ENDFOR
        \STATE $w_\mathcal{G}^{r} \leftarrow$ $\texttt{optimizer}$.\texttt{update}($\eta, \frac{1}{|\mathcal{S}\cup\mathcal{C}|}\sum_{k \in \mathcal{S}\cup\mathcal{C}} \Delta w_{k}^{r})$ 
        \STATE $w_\mathcal{T}^r \leftarrow \begin{cases} \delta w_\mathcal{T}^{r-1} + (1-\delta)w_\mathcal{G}^{r} ~ , r \equiv_u 0\\ w_\mathcal{T}^{r-1} \quad \textrm{else}. \end{cases}$ \COMMENT{EMA update}
		\ENDFOR
	\end{algorithmic}
\end{algorithm}

\noindent Semi-supervised learning approaches typically employ a strong teacher model to machine transcribe audio data, which enables learning in the absence of human labeled supervised data. In compute, communication and memory constrained settings such as on-device federated learning, larger teacher models with higher resource requirements may not be feasible. In this work, we conform with \emph{federated} constraints, and assume that the teacher model is of an equivalent configuration to the student model, can be stored and run on-device, and is used to process audio for machine labeling. 

Algorithm \ref{alg:self-learn} presents the details of the self-learning method. In each training round, we have unlabelled audio from the device for which we obtain the labels using the paired teacher model filtered to exclude utterances of very low or high confidence. Multiple local update steps may be taken on each device (similar to \emph{FedAvg} \cite{mcmahan2017communication}), or a single gradient update step may be taken (similar to \emph{FedSGD}). The gradients are obtained using unlabeled audio on-device, with an augmented form of the audio and the teacher label. The server update step uses the aggregated local model deltas as a pseudo-gradient for its update. Finally, at the end of each training round based on an update frequency, the teacher model is updated using an exponential moving average (EMA \cite{manohar2021kaizen}) of itself and the latest updated global student ASR model. This setup is illustrated in Fig. \ref{fig:self_learning_architecture}.

To help the model mitigate error feedback loops and catastrophic forgetting on older test sets, batches consisting of historical utterances with ground truth transcriptions can be included along with the self-learning updates that use unlabeled data. This process is termed as \textit{rehearsal training}. The rehearsal updates are performed on the cloud by treating the cloud servers as a pseudo-devices and serves as a regularization term to prevent worsening ASR performance. 

\vspace{-5mm}
\subsection{Weak supervision}
\vspace{-2mm}

Weak supervision signals can be used to further improve the performance of the system by leveraging information beyond just the unlabeled audio that self-learning relies on. This work exploits information weaker than the ground truth ASR transcription, which could be recovered from user interactions with the conversational agent. For example, if a user \emph{stops}, \emph{cancels} or \emph{repeats} a request in the subsequent turn of a dialog, it indicates that the previous query was unsuccessfully processed by the device. We study updating ASR models with the help of such a feedback score, potentially indicating whether the user's request was unsuccessful. Further, the correct natural language understanding (NLU) semantics in the form of the correct slot value may eventually be recovered, for e.g., through an explicit re-invocation by the user. Hence, we also study leveraging weak feedback in the form of the NLU slot labels. An example of weak supervision for an utterance can be seen in Table \ref{tab:eg-weak-supervision}. 

\noindent In this work, we demonstrate the impact of weak supervision labels in two forms: (1) machine generated NLU semantics: from an alternate spoken language understanding (SLU) built from ASR$\rightarrow$NLU as a proxy for inferred semantics from user session data; (2) synthetic user feedback scores: a proxy for real user corrections, and available only for the hypothesis served to the user. This framework can accommodate many types of weak supervision information.  

\begin{table}
	\centering
	\caption{Examples of weak supervision available for an utterance. Here, semantic cost (fraction of slots incorrect) is illustrated as the feedback signal.}\label{tab:eg-weak-supervision}
	\begin{tabular}{|p{2.2 cm}|p{5 cm}|}
		\hline
		\textbf{Transcription} & play Halo by Beyonce in main speaker \\
		\hline
		\textbf{ASR hypothesis} & play Hello by Beyond in main speaker \\
		\hline 
		\textbf{NLU semantics} & PlaySong, Artist:Beyonce, Song: Halo, Device: Main speaker \\ 
		\hline
		\textbf{semantic cost} &  2/3 \\
		\hline
	\end{tabular}
	\vspace{2mm}
\end{table}

\subsubsection{Weak Supervision: NLU semantics}
\label{subsec:nlu-semantics}
\vspace{-2mm}

Machine generated NLU semantics from an alternative ASR and NLU model are used as a form of weak NLU feedback, e.g.\ prior work \cite{ponnusamy2020feedback} has used NLU feedback generated by rewriting utterances. Treating the NLU semantics $z$ consisting of the slot type and values from this alternate system as ground truth, we can compute a semantic cost metric $M(z, \textbf{y}_i)$ for an ASR hypothesis. The semantic cost metric is computed for a given hypothesis, as the fraction of slots that have an error. A slot is considered to have an error if the tokens within the slot are not all present in the hypothesis. 
For the purpose of experimentation, we also study the impact of using the alternate system's ASR transcript in addition to the NLU semantics. In this case, the cost $M$ can include the word error rate (WER) obtained comparing $\mathbf{y}_i$ with the alternate transcript $z_t$. For ease of exposition, we consider $z$ to encapsulate both semantics and transcription $z_t$.

\noindent To leverage feedback from these possibly erroneous NLU semantics, we train a model with weight $w$ where the self-learning loss is augmented (summed) with this loss term from the weak NLU signal:
\begin{align}
    &\mathcal{L}_{\textrm{weak}}(w, \mathbf{x}, z) = \E_{y\sim p_w(\mathbf{y}|\mathbf{x})}[M(\mathbf{y}, z)]\nonumber\\
    &\approx \sum_i \hat{p}_w(\mathbf{y}_i|\mathbf{x}) M(\mathbf{y}_i, z) \label{eq:grad-approx}\\
    &\implies \nabla_w \mathcal{L}_{\textrm{weak}}(w, \mathbf{x}, z) \approx \sum_i M(\mathbf{y}_i, z) \nabla_w \hat{p}_w(\mathbf{y}_i|\mathbf{x})\nonumber,
\end{align}
where $\hat{p}_w(\mathbf{y}_i|\mathbf{x}) = p_w(\mathbf{y}_i|\mathbf{x}) / \sum_j p_w(\mathbf{y}_j|\mathbf{x})$ is the normalized probability of the hypothesis. By making an assumption in ~\eqref{eq:grad-approx}, that the probability mass is concentrated in the n-best hypothesis of ASR, the expectation can be approximated by only considering this subset of hypotheses \cite{rao2021mean}. We note that $p_w$ is a differentiable function of $w$ and hence a gradient $\nabla_w \mathcal{L}$ can be computed. 

\vspace{-3mm}
\subsubsection{Weak Supervision: Feedback Scores}
\vspace{-2mm}

In Sec.\ \ref{subsec:nlu-semantics}, we made an assumption that we can obtain weak NLU semantics, and thus have feedback for any hypothesis $\mathbf{y}_i$. Here, we add a constraint that weak supervision is only available for the hypothesis served to the user. The formulation with this constraint, termed weak supervision based on feedback scores, more closely simulate real user feedback where the user has provided feedback only for the served recognition.

\noindent We study two forms of feedback scores - (1) the semantic score as detailed in Sec.~\ref{subsec:nlu-semantics} applied only to the served hypothesis and (2) a binary feedback cost based on the sentence error rate with the true transcription $z_t$, $M(\mathbf{y}, z_t) = \mathbbm{1}(\mathbf{y} \neq z_t)$ (as a proxy for binary user corrections). To simulate an estimation error of the feedback from user interactions, we add a noise term to the feedback signal obtained i.e. $M'(\mathbf{y}, z) = M(\mathbf{y}, z) + U$, with random variable $U$ arising from an arbitrary noise distribution. This helps capture asymmetry and non-uniformity in the feedback from user interactions. 


\noindent The learning is performed with a policy gradient setup. We use the n-best hypotheses to approximate the output lattice/space. A hypothesis ({\em action}) is selected from it by sampling based on the normalized n-best hypotheses probabilities. For the selected hypothesis, we use the feedback $M'(\mathbf{y}, z)$ described above as a {\em reward function} for the policy gradient method to update $w$ which in turn parameterizes $\hat{p}_w(\mathbf{y}_i|\mathbf{x})$. We use the REINFORCE~\cite{williams1992simple,rao2021mean} trick in conjunction with the above to obtain gradients so as to update $w$.
Now,
\begin{align}
    \nabla_w \mathcal{L}_{\textrm{weak}}(w, \mathbf{x}, z) &= \E_{\mathbf{y}\sim p_w(\mathbf{y}|x)}[M(\mathbf{y}, z)\nabla_w\log(p_w(\mathbf{y}|\mathbf{x}))] \nonumber\\
    &\approx M'(\mathbf{y}, z) \nabla_w log(p_w(\mathbf{y}|\mathbf{x})), \mathbf{y}\sim p_w(\cdot|\mathbf{x})\nonumber,
\end{align}
where we take a sampling approximation of size 1 as an estimate of the expectation. With the above setup in place, this framework falls into the premise of Algorithm 1.

\vspace{-3mm}
\section{Experiments}
\vspace{-2mm}

\noindent \textbf{Data}\\
\noindent Our federated continual training experiments are run from January to June $2021$. We use an internal voice-assistant dataset with de-identified utterances totalling $4500$ hours in this time period from $800$K devices. We make only a \emph{single pass} through this data as one of the constraints is that persistent audio storage is not feasible. 

\noindent We evaluate the models on in-house human transcribed (HT) test sets. There is no speaker overlap between the train and evaluation datasets. \textit{General} comprises a $37$-hour test set in $2021$ and older test sets in $2020$. \textit{Delta} comprises a $22$-hour HT test set that records a change in frequency of words in $2021$ over $2020$. The transcriptions are filtered based on $1$-gram, $2$-gram and $3$-grams that are $5$x more frequent in $2021$ than $2020$. This test set captures changes in the data distribution such as new use cases and is crucial to measure the impact of continual learning.

\noindent We also demonstrate results on models trained on public test sets. We use RNN-T models pretrained on the $960$ hour Librispeech dataset \cite{panayotov2015librispeech} and finetuned using self-learning with weak supervision on the 56 hour SLURP dataset \cite{bastianelli2020slurp}. For the public SLURP dataset, we evaluate on the test partition with $13$K utterances. 

\noindent \textbf{Model} \\
\noindent The RNN-T model used contains ~$60$M parameters with a $5\times1024$ LSTM encoder, a $2\times1024$ LSTM prediction network and a feed-forward joint network with {\em tanh} activation~\cite{graves2013speech}. The input embeddings of the prediction network are $512$ dimensional. We use a $2500$ sub-word piece tokenizer \cite{Kudo2018SubwordRI}. The audio features are $64$ dimensional log-mel filter-bank energy features that are computed on a $25$ms window, with a $10$ms shift. SpecAugment~\cite{park2019specaugment} is used for the audio features. The features computed on $3$ consecutive $10$ms frames are stacked and sub-sampled to result in $192$ dimensional features at a $30$ms frame rate, provided as input to the ASR model.

\noindent A $480K$-hour pre-training dataset (where 120K hours are human transcribed and rest machine transcribed) is utilized for pre-training the baseline. Experiments using multiple losses, have equally weighted losses (no tuning). All results shown are using FedSGD with $400$ devices randomly chosen for each of $3000$ training rounds, batch size $16$ and server-side Adam optimizer. For rehearsal training, $40$ cloud pseudo-devices additionally used with historic transcribed data.

\noindent \textbf{Metric} \\ 
\noindent The performance of these models on the voice-assistant data is measured in terms of relative word error rate reduction (WERR) over the initial baseline model at the start of 2021. Positive WERR values represent improvements, while negative ones show degradations. Absolute WER numbers are reported on SLURP experiments.

\section{Results}
\vspace{-3mm}

\renewcommand{\tabcolsep}{2pt}
\begin{table}[t]
	\centering
	\caption{Performance of federated self-learning with weak supervision on the SLURP dataset, including examples of corrected utterances.}\label{tab:res-slurp}
	\begin{tabular}{|p{5 cm}|p{1 cm}|}
		\hline
		\textbf{Setting} & \textbf{WER} \\
		\hline
		Initial & 28.70 \\
		Oracle supervised finetuning & 16.95\\
		\hline
		\multicolumn{2}{|c|}{\textbf{Self-learning}} \\
		\hline
		Teacher not updated & 23.52 \\
	    Teacher updated with EMA & 18.95 \\
		$\quad$+weak-supervision & 18.79 \\
		\hline
	\end{tabular}
	{ \footnotesize 
	\begin{tabular}{|c|p{6 cm}|}
	\hline
	Truth & please help me turn on the robot vacuum cleaner \\
	Initial & please tell me turn on the roblox i can clean \\
	Self-learn & please tell me turn on the robot vacuum cleaner \\
	\hline
	Truth & look for this playback in audiobook and play for me \\
	Initial & look for display light audiobook and play for me \\
	Self-learn & look for this playback in audiobook and play for me \\
	\hline
	Truth & olly what else do i have on the list \\
	Initial & what else do i have in the list \\
	Self-learn & ollie what else do i have on the list \\
	\hline
	\end{tabular}
	\vspace{-3mm}
	}
	\vspace{-5mm}
\end{table}

\begin{table}[t]
	\centering
	\caption{Performance of federated self-learning with weak supervision on voice-assistant data. WERR numbers are relative to WER of the initial model. Multiple forms of weak supervision such as ASR and NLU labels from an alternate SLU model, and NLU feedback scores for the hypothesis served are contrasted.}\label{tab:res-nlu-weak}
	\begin{tabular}{|p{2.8 cm}|p{1.2 cm}|p{1.2 cm}|p{1.2 cm}|}
		\hline
		\textbf{Weak Supervision method} & \textbf{Teacher Update} & \textbf{General WERR} &\textbf{Delta WERR} \\
		\hline
		- & - & -8.16 & -0.02 \\
		- & $\checkmark$ &  -6.12 & 8.29 \\
		 ASR & $\checkmark$ & -1.84 & 11.43 \\
		ASR + NLU & $\checkmark$ &  -1.22 & 11.56 \\
		NLU feedback-score & $\checkmark$& -1.64 & 12.06\\
		\hline
	\end{tabular}
\end{table}

\textbf{Federated self-learning with weak supervision:} We see the performance of self-learning of a pretrained RNN-T model on the public SLURP dataset in Table \ref{tab:res-slurp} that shows self-learning improving the performance by $19\%$ with additional gains from using weak supervision composed of NLU feedback scores. We note that limited gains arise from weak supervision as SLURP has sparse annotations for transcript tokens or few slots per utterance. In few corrected examples, we see self-learning with weak supervision correcting deletion errors and even learning new words like the keyword `olly'. 

\noindent In Table \ref{tab:res-nlu-weak}, the performance of self-learning coupled with weak supervision is depicted for continual learning with a single pass on the internal dataset. First, we observe that if we do not update the paired teacher model with EMA, performance on the new use case does not improve. If we only do self-learning for ASR, there is an improvement of $8.3$\% on the new use case test set. Coupling this with an ASR based weak supervision (where each hypothesis gets a feedback score of the WER computed using a teacher model), we see more improvement that increases as feedback includes the NLU component. We also see similar improvement using only the NLU-based feedback-score obtained only for the served hypothesis as opposed to obtaining a score for all possible hypotheses.

\noindent\textbf{Noisy feedback:}  Table~\ref{tab:synthetic-cpdr} shows the result of federated learning only with noisy feedback for a single served hypothesis from ASR. Here we consider noisy feedback of the form, $M'(\mathbf{y}, z) = M(\mathbf{y}, z) + (-1)^{M(\mathbf{y}, z)}U'$, where random variable $U'\sim p(U| U\in[0,1]), U\sim \mathcal{N}(0, \sigma^2)$ is drawn from a normal random variable with variance $\sigma^2$ truncated to be in the range $[0, 1]$. We then add different levels of noise to measure its impact. In a noisy version of a binary feedback score, 
\begin{align*}
    \E{[M'(\mathbf{y}, z)]} &= \E[M(\mathbf{y}, z)] + \mu \E[(-1)^{M(\mathbf{y}, z)}] \\
    &= (1-2\mu)\E[M(\mathbf{y}, z)] + \mu \\
    \implies \nabla_w \E[M'(\mathbf{y}, z)] &= (1-2\mu) \nabla_w \E[M(\mathbf{y}, z)],
\end{align*}
where $\mu=\mathrm{E}[U']$. Thus if the mean is less than $0.5$, gradient update with the noisy feedback, in expectation, is in the same direction as the gradient update with the true feedback. We demonstrate that even at a high level of noise of $\sigma = 0.4$ we are still able to improve the model on the delta dataset significantly.

\begin{table}[t]
	\centering
	\caption{Performance of learning with only noisy feedback scores on voice-assistant data}\label{tab:synthetic-cpdr}
	\begin{tabular}{|p{5 cm}|c|}
		\hline
		\textbf{Setting}  &\textbf{Delta WERR} \\
		\hline
		binary feedback without noise & 14.45 \\
		binary feedback + noise ($\sigma = 0.1$) & 9.05 \\
		binary feedback + noise ($\sigma = 0.2$) & 7.41 \\
		binary feedback + noise ($\sigma = 0.4$) & 4.40 \\
		\hline
	\end{tabular}
	\vspace{-4mm}
\end{table}

\begin{table}[t]
    \caption{We study (i) the effect of rehearsal training in mitigating the catastrophic forgetting (left) and (ii) the effect of hyper parameters (right) in self-learning on voice-assistant data}\label{tab:ablations}\footnotesize
    \parbox{.45\linewidth}{
    \centering
    \begin{tabular}{|l|p{1cm}|p{1cm}|}
        \hline
         \textbf{Setting} & \textbf{Delta WERR} & \textbf{General (2020) WERR}\\
         \hline
         Self-learning & 14.08 & -13.63 \\
           $\;$+ rehearsal training & 12.47 & -5.85 \\
        \hline
    \end{tabular}
    }
    \hfill
    \parbox{.45\linewidth}{
    \centering
    \begin{tabular}{|l|p{0.9cm}|}
        \hline
         {\em ema} $\delta$, {\em update} $u$  & \textbf{Delta WERR}\\
         \hline
         0.999, 10 & \textbf{14.08} \\
         0.999, 100 & 10.38 \\
         0.999, 200 & 11.56 \\
         0.9999, 10 & 12.64\\
         0.9999, 100 & 11.03 \\
         0.975, 1 & diverge \\
        \hline
    \end{tabular}
   }
\end{table}

\noindent \textbf{EMA hyperparameters and rehearsal training}: In Table~\ref{tab:ablations}, we first see the impact of rehearsal training on mitigating catastrophic forgetting - we observe reduced regression on the older 2020 test set at the expense of performance of new {\em Delta} test sets. Delta test set results are not comparable across prior tables as amount of computation, catastrophic forgetting differ. We also study the impact of EMA hyperparameters, higher $\delta$ implies lower weight to new updates and update frequency $u$ determines how often the teacher model is updated. Improved performance is seen for frequent updates with a lower EMA value. We also observed training diverging when the teacher model is updated to the student model after each step, suggesting that an error feedback loop takes place.


\vspace{-4mm}
\section{Conclusion}
\vspace{-3mm}
We focused on the federated continual learning problem for ASR where an ASR model deployed on-device is updated ensuring that (1) human ground-truth transcriptions are not available, (2) large device compute and memory are not required to run strong teacher models for labelling the audio (3) audio is not persisted or sent to the cloud. We demonstrated that using a paired teacher model to generate labels for the unlabelled audio and where the teacher model is updated using an exponential moving average of the RNN-T model can improve RNN-T performance by $10\%$ on new use cases with larger improvement on public SLURP dataset and only $10\%$ away from the fully supervised setting. Rehearsal training using historical datasets with ground-truth transcriptions mitigates catastrophic forgetting and error feedback loops. We made use of weak supervision signals such as machine generated NLU semantics or simulated noisy feedback scores from interactions of a user in a policy-gradient approach which further improved the performance of self-learning. 


\noindent\footnotesize{\textbf{Acknowledgments:} We thank Gurpreet, Aaron, Buddha, Bach, Harish, Ehry and Shehzad for helpful discussions.}

\bibliography{mybib}
\bibliographystyle{IEEEtran}
\end{document}